\documentclass[
,showpacs
]{revtex4}
\usepackage{epsfig}
\begin{document}
\title{Buckling transition in icosahedral shells subjected to volume 
conservation constraint and pressure: relations to virus maturation}
\author{Antonio \v{S}iber}
\email{asiber@ifs.hr}
\affiliation{Institute of Physics, P.O. Box 304, 10001 Zagreb, Croatia}

\begin{abstract}
Minimal energy shapes of closed, elastic shells with twelve pentagonal disclinations 
introduced in otherwise hexagonally coordinated crystalline lattice are studied. The 
geometry and the total energy of shells are studied as a function of the elastic 
properties of the material they are made of. Particular emphasis is put on the 
{\em buckling transition} of the shells, that is a strong preference of the 
shell shapes to 'buckle out' in spatial regions close to the pentagonal disclinations 
for certain range of the elastic parameters 
of the problem. The transition effectively increases the mean square aspherity of 
shapes, making them look more like an icosahedron, rather than a sphere which is 
a preferred shape prior to the onset of the transition. 
The properties of the buckling transition are studied in cases 
when {\em (i)} the total volume enclosed by the elastic shell has to be fixed and when 
{\em (ii)} there is an internal pressure acting on the shell. This may be related 
to maturation process in non-enveloped dsDNA viruses, where the insertion of the 
genetic material in a pre-formed protein shell (viral coating) may effectively 
impose the fixed volume/pressure constraint. Several scenarios that may explain the 
experimentally observed feature of mature viruses being more 
aspherical (facetted) from their immature precursors are discussed and new 
predictions for the elastic properties of viral coatings are obtained on the 
basis of the presented studies.
\end{abstract}
\maketitle

\begin{center}
Published in Phys. Rev. E {\bf 73}, 061915 (2006)
\end{center}

\section{Introduction}
\label{sec:intro}
Several articles have appeared recently that aim to describe the virus structure, shape and 
stability by using physical principles. Virus related research seems to be particularly 
appealing to physicists since viruses exhibit many features that are reminiscent of 
well known and thoroughly studied phenomena in ''more traditional'' physical systems. For 
example, fifty years ago Fraenkel-Conrat demonstrated that infectious viral particles 
(tobacco mosaic viruses, TMV) can be reassembled from two solutions, one containing the viral genetic material (RNA) and the other its coat (or {\em capsid}) proteins (a short historical 
overview of research on reconstitution of TMV can be found in 
Ref. [\onlinecite{Fraenkel}]). The reassembly of TMV proceeds without any special 
external impetus - it is spontaneous. The fact that viruses can be reassembled 
in {\em in vitro} conditions suggests that their shape and symmetry should be a 
result of free energy minimization, a concept familiar from equilibrium 
thermodynamics. This line of thought has recently been successfully applied in 
the explanation of physical {\em origin} of the icosahedral symmetry of viruses 
(see below) and other possible shapes and geometries that might be adopted by viruses 
\cite{BruinPRL,coarse}. Production of 
functional molecular machine from its constituent, possibly engineered 
molecular components is of course a dream of nanotechnologists. Such approach to 
''molecular engineering'' does not require precise molecular positioning, 
since correct assembly takes place {\em spontaneously} due to specific 
architecture of the molecular constituents and pronounced anisotropy 
of their mutual interactions, hence the term {\em self-assembly} is often 
used in nanotechnological context. 

\begin{figure}[ht]
\centerline{
\epsfig {file=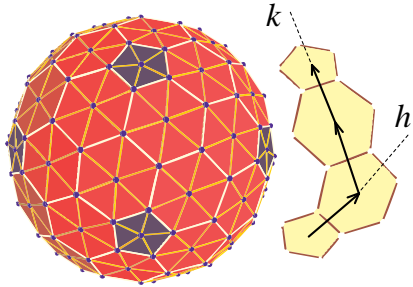,width=7cm}
}
\caption{(Color online) Idealized $T=7$ ($h=1$, $k=2$) virus structure. The structure 
consists of twelve pentamers (at the vertices of an icosahedron) and 
sixty hexamers. Note how the neighboring pentamers 
can be connected with two subsequent translations along two 
distinct spherical geodesics.
}
\label{fig:fig1}
\end{figure}

The structural symmetry and 
shape of virus coating is in itself intriguing. There have been attempts to 
draw parallels between the symmetry of viruses, Penrose tilings \cite{Twarock}, 
and quasicrystals \cite{Casparquasi}. Most viruses have a structure 
with topology equal to that of triangulated icosahedra (see Fig \ref{fig:fig1}).
Individual proteins that make the viral coating (approximated by triangles in Fig. 
\ref{fig:fig1}) are organized in units called 
{\em capsomers} that consist of five (pentamers) or six protein units 
(hexamers). Most viruses contain only twelve pentamers and remaining 
capsomers are hexamers \cite{sv40}. Pentamers are located at twelve vertices 
of an icosahedron. Such structures are very similar to giant 
fullerene molecules \cite{Fulleren1}. It is also easy to see a 
structural relationship between the virus and an icosahedral 
geodesic dome \cite{Geodome}. One of the simplest virus structures 
($T=3$, see below), characteristic 
of small viruses (e.g. cucumber mosaic virus \cite{Viper}), 
contains twelve pentamers and twenty hexamers and 
is topologically equivalent to a famous buckminsterfullerene 
molecule (C$_{60}$) \cite{Bucky}. Icosahedral 
symmetry that is characteristic of viral shapes is also frequently found in 
ground state configurations of atomic clusters (see e.g. 
\cite{Siberclust}). Poligonalizaton of a sphere 
with pentagons and hexagons is often described in terms of Caspar-Klug 
quasiequivalent construction \cite{Caspar-Klug}. Within this framework, 
viruses are characterized by the so-called {\em $T$ number} which describes 
the order of a poligonalization. The $T$ number can be written as 
$T=h^2 + k^2 + hk$, where $h$ and $k$ are nonnegative integers. The 
meaning of these integers can be quickly grasped from Fig. \ref{fig:fig1}. 
Note how the two neighboring pentamers in the structure displayed 
in Fig. \ref{fig:fig1} can be connected through two subsequent 
translations along two distinct spherical geodesics. The magnitudes of these 
translations are nonnegative integer multiples ($h$ and $k$) of distances between 
the nearest-neighboring capsomers, so as that the structure in Fig. \ref{fig:fig1} 
can be characterized by $h=1$ and $k=2$, or $T=7$. The total number of capsomeric units in 
a virus of particular $T$-number symmetry is $10T + 2$. Twelve of 
these units are pentamers and remaining $10(T-1)$ are hexamers \cite{sv40}.

The shapes of viral capsids have been recently studied within the 
framework of nonlinear theory of elastic shells \cite{elastic1,elastic2}. These studies 
have produced rather interesting results and excellent fits to experimentally 
determined virus 
shapes have been obtained \cite{elastic1}. The most important feature of 
the shell shapes that these studies predict is the {\em buckling transition}. 
The buckling transition has been earlier predicted for planar triangular meshes with 
a pentagonal or heptagonal defect introduced in their structure \cite{Seung}. 
As a consequence of minimization of the total energy, 
which includes both stretching and bending contributions, the crystalline mesh buckles 
in a conical shape in the vicinity of a pentagonal defect (or disclination). This transition 
is observed only above the critical value of the so-called 
{\em Foppl-von K\'{a}rm\'{a}n number} ($\gamma$) \cite{Seung}, which is an effective 
parameter combining the mesh radius, two-dimensional Young's modulus ($Y$) and 
bending rigidity ($\kappa$). This parameter uniquely describes the shape of continuous shells. 
A pentagonal ''defect'' (pentamer) is characteristic of a virus 
structure, and viral coating can thus be viewed as a continuous shell with twelve 
pentagonal disclinations introduced in otherwise hexagonally 
coordinated crystalline structure (each vertex in the ''regular'' mesh has 
six nearest neighbors). The 
two studies \cite{elastic1,elastic2} have reached the conclusion that the remnants 
of buckling transitions survive in the more complex geometry of a 
spherical (closed) shell with twelve pentagonal disclinations. This transition is not 
as sharp as in the case of a (two-dimensional) disk, nevertheless an observable 
buckling transition of the shell was found for F\"{o}ppl-von K\'{a}rm\'{a}n (FvK) 
numbers between about $\gamma=100$ and $\gamma=1000$. This effect has been related 
to the features of the experimentally determined virus shapes, and it 
can correctly describe the global trend of larger viruses being more faceted (or 
buckled) than smaller ones \cite{Review3Dviruses}. One of the interesting 
propositions put forth by the authors of Ref. \cite{elastic1} is that the 
process of virus maturation can be described in part by the buckling 
transition of viral capsids. Virus maturation is the process through which the 
assembled viral particles become fully functional. Its precise evolution 
depends on the type (or family) of the virus in question, in particular whether the viruses are 
enveloped by the cell membrane proteins or not. For a large class of non-enveloped viruses 
[e.g. those containing the double stranded DNA (dsDNA) in their mature 
form \cite{Review3Dviruses}], 
the maturation consists of incorporation of the 
genetic material in a preassembled, empty protein capsid (the so called precursor 
capsid or {\em procapsid}). During the maturation process the capsid typically 
''swells'' and mature capsids are larger from their precursor counterparts.
A faceted shape is often a characteristic of a mature virus, whereas (immature) 
precursor capsids have a more spherical shape. This transition in shape has been 
described within a framework of the elastic theory of shells in terms of the change of the 
FvK number \cite{elastic1} - the FvK number in mature viral shells is larger than 
in their respective precursor capsids. However, during the buckling transition 
of elastic shells, the volume enclosed by the shell {\em decreases}, whereas 
for most viruses, the mature viral capsids enclose {\em larger} volume from the 
immature precursor capsids \cite{Conway,Jiang}.

The aim of this article is to investigate whether the buckling transition 
in elastic shells survives in more restrictive circumstances, in particular under 
the constraint of volume conservation and when the internal pressure acting 
on the shell is nonvanishing. These constraints may be thought of as the 
simplest possible introduction of the capsid-DNA/RNA interaction in the 
problem of virus shapes. Thus, an investigation of the 
characteristics of buckling transition in more general conditions 
should be of use for a more thorough understanding of viral capsid shapes in their 
immature and mature forms.
 
The article is organized as follows. Section \ref{sec:model} briefly discusses the 
model of elasticity and minimal energy shapes of shells with 12 pentagonal 
disclinations subjected to a constraint of fixed volume. 
Section \ref{sec:empty} contains the results pertaining to empty shells, 
i.e. without the imposition 
of the fixed enclosed volume or constant pressure constraints. 
This section deals with the subject already 
treated in Refs. [\onlinecite{elastic1}] and [\onlinecite{elastic2}], and 
its main purpose is to clearly demonstrate that the {\em shell volume 
decreases during the buckling transition}.
In section \ref{sec:full}, the minimal energy shell shapes with the imposed 
constraint of fixed enclosed volume are studied. It is shown that 
the buckling transition 
survives in that case also, although the buckled shapes are less aspherical than the 
ones obtained without the volume conservation constraint. 
Section \ref{sec:press} contains results regarding the buckling transition 
of shells subjected to constant environmental pressure (the 
difference in pressures in the inner and outer space of the shell is 
fixed). Section \ref{sec:appli} relates the results of the article to the virus 
maturation process. Several possible scenarios that reproduce the 
experimentally observed fact of mature viruses being more aspherical, facetted 
from their immature precursors are discussed. New predictions for 
the parameters characterizing the elastic response of viral coatings are 
given. Section \ref{sec:discuss} briefly 
summarizes the results and concludes the article.

\section{A model of elastic shells with the constraints of fixed enclosed volume 
and constant pressure} 
\label{sec:model}

I start from the model of elasticity described in Refs. 
[\onlinecite{Seung,elastic1,elastic2}]. Briefly, the shell surface 
is discretized in triangular plaquettes and the Hamiltonian describing 
such system is
\begin{equation}
H = \frac{\epsilon}{2} \sum_{i,j} (|{\bf r}_i - {\bf r}_j| - a)^2 
+ \frac{\tilde{\kappa}}{2} \sum_{I,J} ({\bf n}_I - {\bf n}_J)^2, 
\label{eq:h0}
\end{equation}
where indices $i$ and $j$ ($I$ and $J$) describe the two neighboring 
triangle vertices (surfaces) located at ${\bf r}_i$ and ${\bf r}_j$ 
respectively, $\epsilon$ is the scale of energy related 
to a change of the distance between the two neighboring vertices (stretching), while 
$\tilde{\kappa}$ is the scale of energy related to a change of dihedral 
angle (bending) between the two neighboring triangles (those sharing a side) whose 
normal vectors are denoted by ${\bf n}$. The equilibrium distance between 
the two neighboring vertices is $a$, and it is assumed that the neighboring 
triangular plaquettes prefer to lie in the same plane, i.e. the preferred 
angle between their normal vectors is zero. Authors of Ref. 
[\onlinecite{elastic2}] considered also the case of nonzero preferred 
curvature of the surface which can be simply introduced in the 
above Hamiltonian \cite{elastic2}. For surfaces 
containing a very large number of triangular plaquettes, the discrete 
Hamiltonian becomes reliable for the description of continuous elastic medium 
described by Young's modulus $Y=2 \epsilon / \sqrt{3}$, Poisson's ratio 
$\nu = 1/3$, bending rigidity $\kappa = \sqrt{3} \tilde{\kappa} / 2$, and 
Gaussian rigidity $\kappa_G = -4 / 3$ \cite{elastic1}. 

Finding a minimum-energy state (or a shape) of a problem defined by 
Eq. (\ref{eq:h0}) consists of multidimensional search for a minimum 
of the Hamiltonian function $H({\bf r}_1,...,{\bf r}_N)$ that depends on $3N$ 
coordinates, where $N$ is the number of mesh vertices. This search is 
{\em unconstrained} which means that there are no relationships between 
the variables ${\bf r}_1,...,{\bf r}_N$ that have to be obeyed. 

\subsection{Enclosed volume constraint}
It is possible to do a constrained search for the minimum of Eq. (\ref{eq:h0}), 
specifying additionally some relations that the variables (or a shape) have 
to obey. The volume of a shell can be expressed in terms of the coordinates of the 
mesh vertices as
\begin{equation}
V = \frac{1}{6} \sum_I |{\bf r}_{I,1} \cdot ({\bf r}_{I,2} \times {\bf r}_{I,3} )|,
\label{eq:volume}
\end{equation}
where ${\bf r}_{I,1}, {\bf r}_{I,2}$, and ${\bf r}_{I,3}$ are the three vertices 
of the triangle $I$ in (counter)clockwise order. A search for a 
minimum of Eq. (\ref{eq:h0}) with the independent variables (vertex coordinates) 
additionally fulfilling $V = V_0$, where $V_0$ is a fixed quantity, should 
result in a shape that minimizes the function $H({\bf r}_1,...,{\bf r}_N)$ in the 
subspace of all possible shapes that enclose volume $V_0$. The 
volume constraint is implemented in the numerical procedure using the 
so-called penalized version of the Hamiltonian \cite{penalization} where 
an additional, quadratic penalty term of the form
\begin{equation}
H_V = \frac{\lambda}{2} (V - V_0)^2
\end{equation}
is added to the original Hamiltonian in Eq. (\ref{eq:h0}), so that 
the new Hamiltonian ($H_n$) is given by $H_n = H + H_V$. Ideally, penalty 
parameter $\lambda$ should be extremely large (and positive) to strictly 
impose the volume constraint. 
The problem of finding a minimum of a strongly constrained hamiltonian is 
solved using the so-called continuation technique \cite{penalization}. 
A small value of $\lambda$ is chosen initially, which enables a quick numerical 
convergence of the shape. This shape is used as an initial guess for the 
hamiltonian in which $\lambda$ is ten times larger ($\lambda \rightarrow 10 \lambda$), and 
the procedure is repeated until $\lambda$ is large enough so that the 
volume constraint is fulfilled to a machine precision.

\subsection{Constant internal pressure}
\label{sub:tlakham}
The constant internal pressure ($p$) is introduced in the problem by adding 
to the shell Hamiltonian a term $pV$, so that the new Hamiltonian of the 
problem ($H_n$) is given by
\begin{equation}
H_n = H + pV.
\label{eq:tlak}
\end{equation}
This introduces an additional force $\Delta {\bf F}_i$ on the $i$-th vertex 
which is given by $\Delta {\bf F}_i = - p \partial V / \partial {\bf r}_i$.

\section{Empty icosahedral shells}
\label{sec:empty}

A word of caution concerning the $T$-numbers of triangulated shells used 
by the authors of Refs. \cite{elastic1,elastic2} and Caspar-Klug $T$-numbers 
pertaining to viral shells is in order here. It should be understood that these 
two are not the same. For example, for the idealized virus structure in 
Fig. \ref{fig:fig1}, the Caspar-Klug $T$-number is 7 ($h=1, k=2$), while the 
$T$-number characteristic of a kind of triangulation (without reference to 
pentamer/hexamer structures \cite{elastic1,elastic2}) is 21 ($h=4, k=1$, this 
can be obtained by simply following a path from one pentamer to its neighboring 
pentamer with 
integer steps along the sides of the mesh triangles and not through the 
centers of pentamer/hexamer structures). Of course, the two $T$-numbers 
become the same if {\em every} vertex in the triangular mesh corresponds to the 
center of a capsomer. In what follows, I shall 
reserve the term "$T$-number" for the characterization of the shell triangulation, 
while the term "Caspar-Klug $T$-number" shall be used for the characterization 
of $T$-numbers in the context in which they were originally introduced.

\begin{figure}[ht]
\centerline{
\epsfig {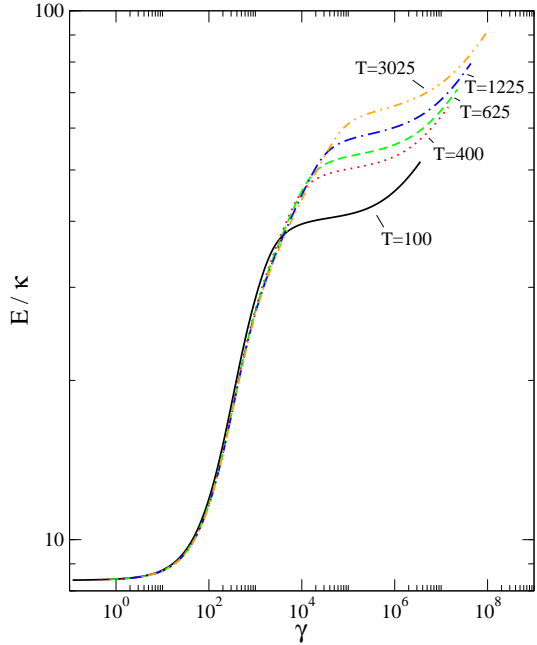}
}
\caption{(Color online) Shell energy as a function of F\"{o}ppl-von K\'{a}rm\'{a}n 
number for several $T$-numbers as denoted in the figure ($T=100,400,625,1225,3025$).}
\label{fig:fig2}
\end{figure}

Icosahedral shells with various $T$-numbers have been constructed and their total 
energy ($E$, which includes both stretching and bending contributions) 
as a function of bending modulus, $\kappa$ has been calculated. Other parameters 
of the Hamiltonian in Eq. (\ref{eq:h0}) have been kept fixed ($a=1$, $\epsilon=1$). 
The initial shape for $\kappa \rightarrow \infty$ has been setup 
[typically a sphere with a radius of 
$a \sqrt{5T\sqrt{3}/\pi} / 2$, see Eq. (\ref{eq:volestimate})], and the minimum energy 
shape was obtained by an efficient implementation of the 
conjugate gradient method described in Ref. [\onlinecite{CGmetoda}]. The resulting 
shape is used as an initial guess for the situation in which $\kappa$ is decreased, and 
this procedure is repeated until $\kappa$ becomes rather small. As demonstrated 
by the authors of Ref. [\onlinecite{elastic1}], the shape and energy 
of continuous shells (large $T$-numbers) are unique functions of the 
FvK number which can be expressed as 
\begin{equation}
\gamma = \langle R \rangle ^2 \frac{Y}{\kappa}, 
\label{eq:foppl}
\end{equation}
where $\langle R \rangle$ is the mean radius of the shell given as 
\begin{equation}
\langle R \rangle = \frac{1}{N} \sum_{i=1}^{N} |{\bf r}_i - {\bf r}_0|, 
\end{equation}
and ${\bf r}_0$ is a geometrical center of the shape, 
${\bf r}_0 = \sum_{i=1}^{N} {\bf r}_i / N$. The total shell energy as a function 
of FvK number is shown in Fig. \ref{fig:fig2} for shells with 
different $T$-numbers.

It can be seen that the continuum regime concerning the total energy of the shell 
is reached rather slowly and large $T$-numbers are needed in this respect. This is 
especially true in the region of large FvK numbers 
($\gamma > 20 000$). Shells with smaller $T$-numbers can reliably predict 
the continuum shapes in the region $0 < \gamma < 20 000$.

\begin{figure}[ht]
\centerline{
\epsfig {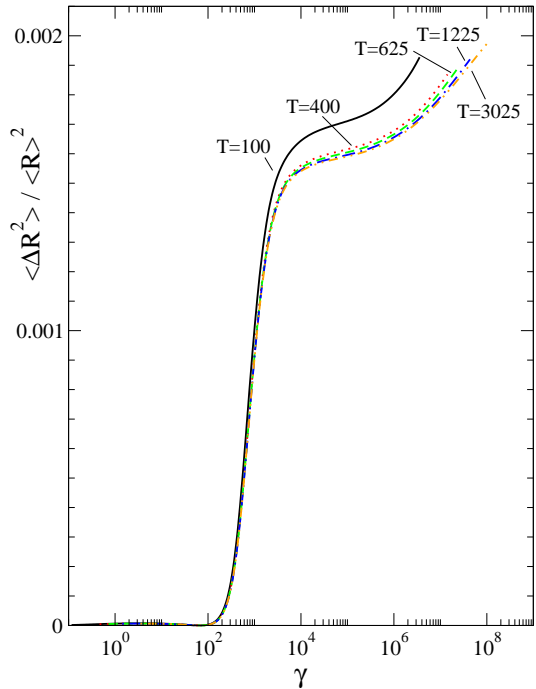}
}
\caption{(Color online) Mean square aspherity of minimal energy shapes as a function of 
Foppl-von K\'{a}rm\'{a}n number for several 
$T$-numbers as denoted in the figure ($T=100,400,625,1225,3025$).}
\label{fig:fig3}
\end{figure}

Based on quite general considerations, 
the authors of Ref. [\onlinecite{elastic1}] concluded that the energy 
of a closed triangular shell (without the enclosed volume conservation constraint) 
with twelve pentagonal disclinations situated at the icosahedron vertices should behave as
\begin{equation}
\frac{E}{\kappa} = \left \{ 
\matrix{
6 B \gamma / \gamma_b + D, \, \gamma < \gamma_b\cr
\cr
6 B [1 + \ln ( \gamma / \gamma_b ) ] + D, \, \gamma > \gamma_b,
}
\right.
\label{eq:forma}
\end{equation}
where $D$ is a constant contribution to the energy due to the "background curvature" of 
a sphere, $B$ is a numerical constant which could be interpreted as a pentagonal 
disclination "core energy" and which should be reasonably close to $\pi / 3$, and 
$\gamma_b$ is a critical FvK number indicating approximately 
the region in which the buckling transition takes place. For the (unbuckled) shapes 
described by $\gamma < \gamma_b$, the total energy is approximately given by 
the sum of elastic stretching energy of twelve ''unbuckled'' pentagonal disclinations
($6 B \gamma / \gamma_b$) and by the ''background'' elastic energy, $D$ - there is 
no buckling contribution related to pentagonal disclinations other than the 
overall, mean background curvature. For the 
buckled shapes ($\gamma > \gamma_b$), the functional dependence of the energy related 
to disclinations profoundly changes, and one has to account for the 
energies resulting from the bending of the conical section in the vicinity of 
each of the disclinations. This results in a logarithmic dependence of energy 
on $\gamma$. Further details can be found in Refs. [\onlinecite{Seung,elastic1,elastic2}].
The data for the largest $T$-number ($T=3025; h=55, k=0$) were fitted to the 
analytical forms in 
Eq. (\ref{eq:forma}), with $B, D$ and $\gamma_b$ treated as fit parameters. I have 
found that the best fit to the numerical results in the region of 
$\gamma \in \langle 0,31000]$ is obtained with $B=1.29$, $\gamma_b = 264$, and 
$D=8.51$. These numbers are in good agreement with the results reported 
in Ref. [\onlinecite{elastic2}] ($B=1.27$, $\gamma_b = 260$ \cite{primjedba}), 
but $\gamma_b$ is two times larger from the value found in Ref. [\onlinecite{elastic1}]. 
One should note, however, that this value is not 
really ''critical'', since none of the characteristics of the shape undergo 
a discontinuous change at $\gamma_b$ (see Fig. \ref{fig:fig3}).

Mean square aspherities of shapes defined as in Ref. \onlinecite{elastic1}, 
\begin{equation}
\frac{\langle \Delta R^2 \rangle}{\langle R \rangle ^2} = \frac{1}{N} \sum_{i=1}^{N} 
\frac{(|{\bf r}_i - {\bf r}_0| - \langle R \rangle)^2}{\langle R \rangle ^2}
\end{equation}
are plotted in Fig. \ref{fig:fig3} as a function of $\gamma$ for the same 
choice of $T$-numbers as in Fig. \ref{fig:fig2}.

The shape geometries apparently converge with the $T$-number faster than the 
total energies (compare Figs. \ref{fig:fig2} and \ref{fig:fig3}). One should 
note how the buckling transition directly transcribes into increased mean 
square aspherities of the minimal energy shapes. The results are 
very similar to those presented in Ref. [\onlinecite{elastic1}].

\begin{figure}[ht]
\centerline{
\epsfig {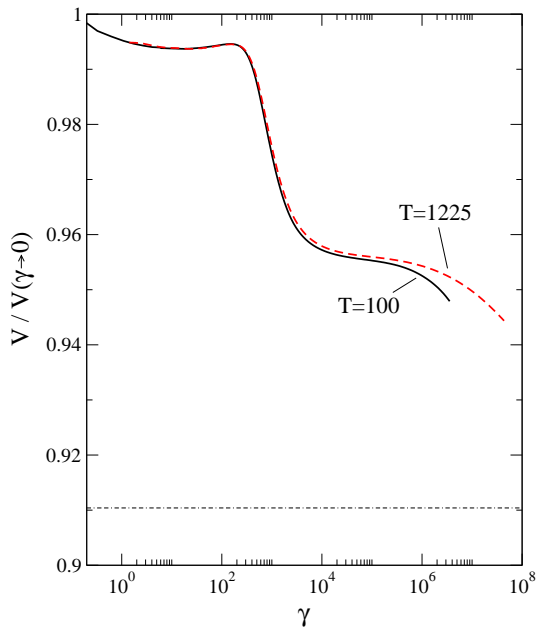}
}
\caption{(Color online) Volume enclosed by the shell as a function of 
F\"{o}ppl-von K\'{a}rm\'{a}n number for 
two values of the $T$-number as denoted in the figure ($T=100, 1225$)}
\label{fig:fig4}
\end{figure}

It is of interest to examine how the volume enclosed by the shell changes during the 
buckling transition. Rough estimates can be easily obtained. For small 
FvK numbers, the minimal energy shapes are very nearly 
spherical, while for extremely large FvK numbers, the 
shapes are nearly 
perfect icosahedra. Assuming that the shape area is the same in both limits, 
the volume ratio should be close to 
\begin{equation}
\frac{V (\gamma \rightarrow \infty)}{V (\gamma \rightarrow 0)} \approx 0.91
\label{eq:omjervol}
\end{equation}
In fact, $V (\gamma \rightarrow 0)$ can be well approximated as 
\begin{equation}
V (\gamma \rightarrow 0) \approx \frac{a^3}{6 \sqrt{\pi}} 
\left( 5 T \sqrt{3} \right)^{\frac{3}{2}}.
\label{eq:volestimate}
\end{equation}
The above expression neglects the elastic strain in the spherical structure and 
assumes that all the triangular faces of the mesh have equal areas of 
$a^2 \sqrt{3} / 4$, which is of course not true, especially in the regions 
close to pentagonal disclinations. Nevertheless, the numerically calculated 
value of $V (\gamma \rightarrow 0)$ was found to be only about 0.7 \% smaller 
from the estimate in Eq. (\ref{eq:volestimate}).
Calculated changes of volume during buckling transitions are presented in 
Fig. \ref{fig:fig4}. It can be seen that only about half of the maximum volume 
reduction predicted by Eq. (\ref{eq:omjervol}) takes place during the buckling transition 
(for $100<\gamma<2000$), while the remaining part of the reduction takes place 
in the icosahedron's ridge sharpening regime (for $\gamma > 10^6$) studied in 
Refs. [\onlinecite{Lobkovsky1,Lobkovsky2}] and discussed also in 
Refs. [\onlinecite{elastic1,elastic2}].

\section{Buckling transition in icosahedral shells subjected to the fixed enclosed volume constraint}
\label{sec:full}

The ''model'' physical system that should be helpful in comprehending the 
results of this section is that of a formed shell 
with $\kappa \rightarrow \infty, \gamma \rightarrow 0$ (a sphere) into which an 
incompressible liquid is poured. The bending modulus of the shell is 
decreased and the change in energy and shape are monitored during the process. This 
picture is useful when the constrained volume is larger than the one in 
the limit $\kappa \rightarrow \infty, \gamma \rightarrow 0$, and the incompressible 
liquid is a rough approximation of the viral genetic material.

\begin{figure}[ht]
\centerline{
\epsfig {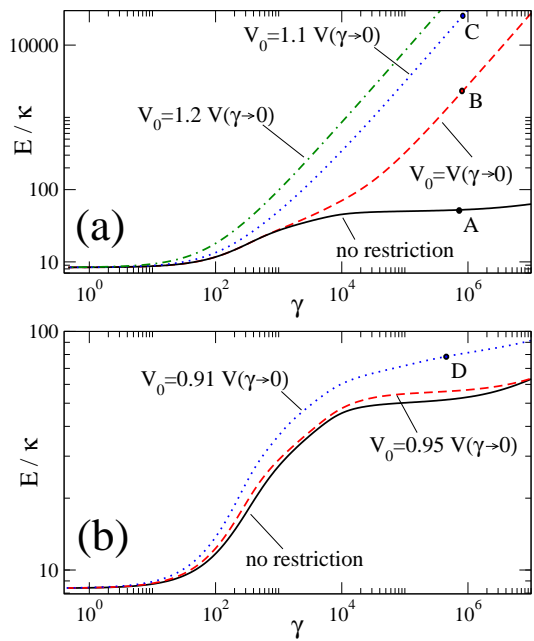}
}
\caption{(Color online) Comparison of shell energies with and without the 
volume conservation constraint as a function of F\"{o}ppl-von K\'{a}rm\'{a}n 
number for $T=400$. Panel (a): Enclosed volumes are constrained to values 
larger (or equal) than $V(\gamma \rightarrow 0)$; $V = V_0 = V(\gamma \rightarrow 0)$, 
$V = V_0 = 1.1 V(\gamma \rightarrow 0)$, and $V = V_0 = 1.2 V(\gamma \rightarrow 0)$ 
as denoted in the figure. 
Panel (b): Enclosed volumes are constrained to values 
smaller than $V(\gamma \rightarrow 0)$; $V = V_0 = 0.95 V(\gamma \rightarrow 0)$ and 
$V = V_0 = 0.91 V(\gamma \rightarrow 0)$ as denoted in the figure. The energy of 
unconstrained shells is also shown for comparison.}
\label{fig:fig5}
\end{figure}

For the constrained shapes, FvK number is again calculated as in 
Eq. (\ref{eq:foppl}), although its meaning as the {\em sole} parameter that uniquely 
describes the shell shape \cite{elastic1} is obviously lost. The results of 
the calculation of shell energies with and without the enclosed volume constraint are 
presented in Fig. \ref{fig:fig5}. 
Volumes $V$ that are larger than $V(\gamma \rightarrow 0)$ seem to be of interest 
to virus maturation process. To complete the picture of buckling transition in 
elastic shells (without respect to viruses), I have also considered the constraints to the 
total volume that are smaller than $V(\gamma \rightarrow 0)$. 

The energies presented in Fig. \ref{fig:fig5} correspond only to the energy contained 
by the shell. For more realistic application of a shell theory to the viral shapes, one 
should also include the contribution of the capsid - DNA/RNA interaction in the total 
energy. It can be easily observed from panel (a) of Fig. \ref{fig:fig5} that the 
aspherical 
shell (viral) shapes become energetically very expensive for FvK numbers larger 
than about 10$^4$ and it is thus unlikely that shapes characterized by such large 
FvK numbers will be adopted by mature viral shells. This statement 
is corroborated 
by the fits of structures of bacteriophage HK97 and L-A yeast virus to the shapes 
predicted by Hamiltonian in Eq. (\ref{eq:h0}) which produced values 
of $\gamma = 1480$ and $\gamma=547$, respectively, \cite{elastic1} which are in 
the region of $\kappa$'s where 
the volume conservation constraint does not increase the shell energy significantly 
(see Fig. \ref{fig:fig5}). A critique and limitations of this line of thinking and 
other possible explanations of the buckling transitions are presented in 
Sec. \ref{sec:appli}. 

Note that the shell energies for volume constraints $V = V_0 < V(\gamma \rightarrow 0)$ 
(panel (b) of Fig. \ref{fig:fig5}) do not increase significantly with above 
those obtained without the volume constraint (compare energy scales in 
panels (a) and (b) of Fig. \ref{fig:fig5}). Note that the volume constraint 
of $V = V_0 = 0.95 V(\gamma \rightarrow 0) \approx V(\gamma = 10^7)$ is {\em not} a 
constraint at the point when $\gamma \approx 10^7$, since at that point the volume 
adopted by the 
unconstrained shell equals to the constrained volume (see Fig. \ref{fig:fig4}). 
Thus, the energy curves for the unconstrained shape 
and the shape constrained to $V = V_0 = 0.95 V(\gamma \rightarrow 0)$ 
touch at $\gamma \approx 10^7$.

\begin{figure}[ht]
\centerline{
\epsfig {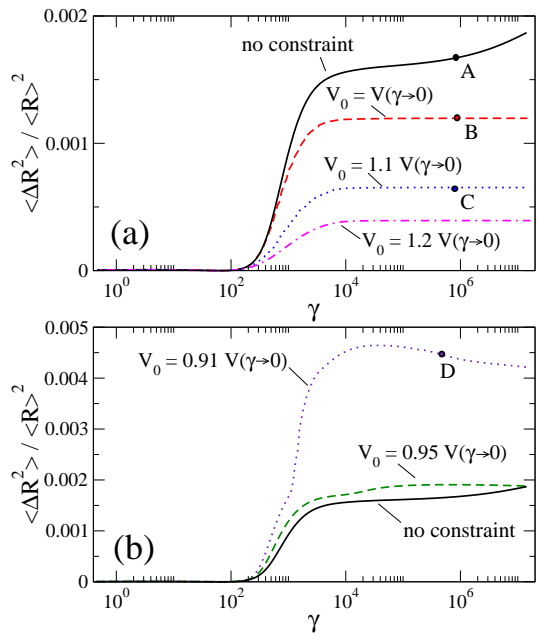}
}
\caption{(Color online) Mean square aspherity of unconstrained and constrained shells as a 
function of the F\"{o}ppl-von K\'{a}rm\'{a}n number for $T=400$. The 
volume constraints studied are the same as in Fig. \ref{fig:fig5}, and 
panels (a) and (b) display the results for constraints 
$V = V_0 \ge V(\gamma \rightarrow 0)$ and $V = V_0 < V(\gamma \rightarrow 0)$, respectively.}
\label{fig:fig6}
\end{figure}

\begin{figure}[ht]
\centerline{
\epsfig {file=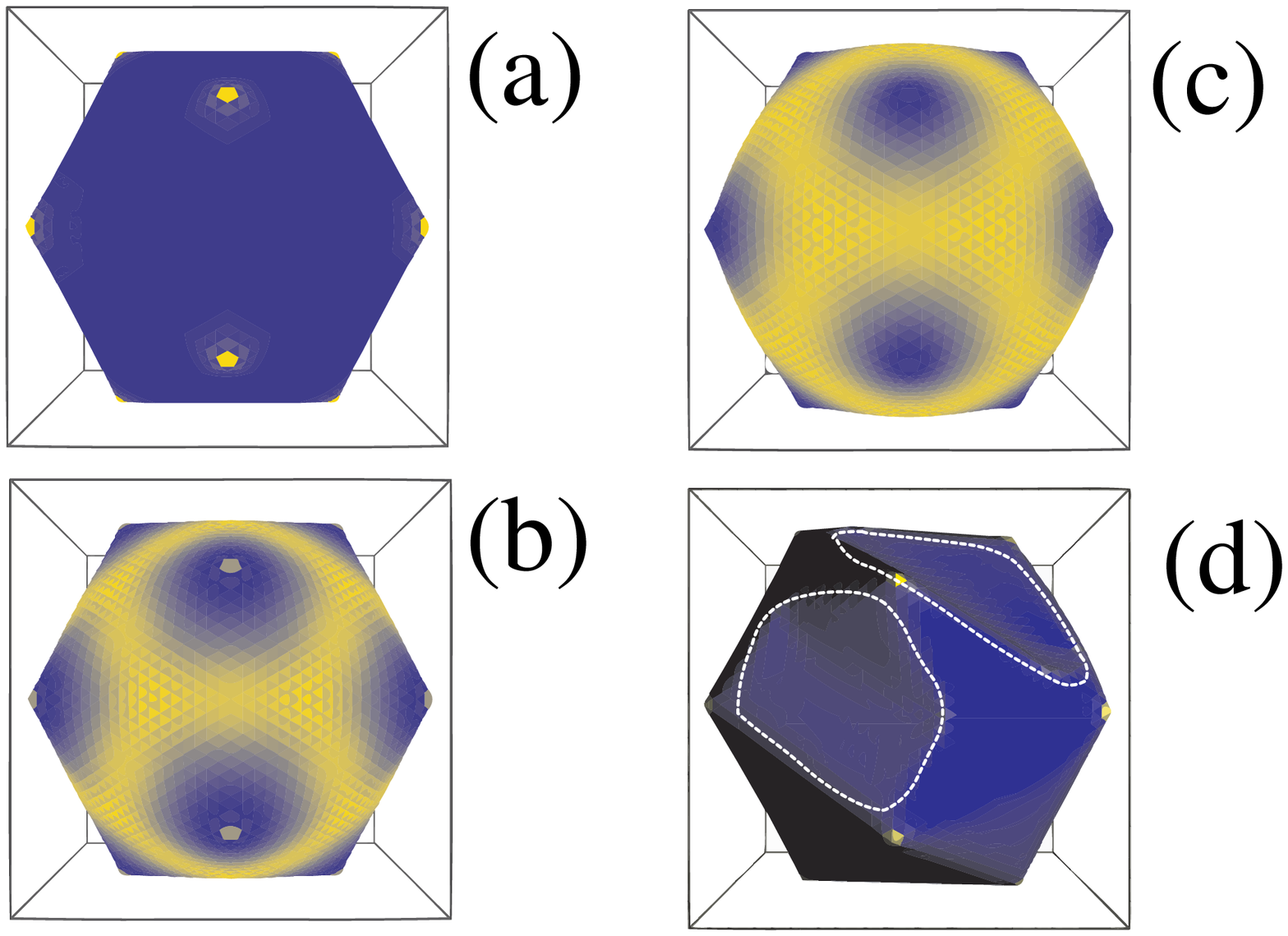,width=7cm}
}
\caption{(Color online) The buckled shapes for the Hamiltonian parameters denoted by A 
[panel (a); no volume constraint), B [panel (b); $V = V_0 = V(\gamma \rightarrow 0)$], c) 
[panel (c); $V = V_0 = 1.1 V(\gamma \rightarrow 0)$] and D 
[panel (d); $V = V_0 = 0.91 V(\gamma \rightarrow 0)$] in Figs. \ref{fig:fig5} and 
\ref{fig:fig6}. The faces are colored according to their total energy. 
The brighter (yellow) triangles correspond to faces with large energy, while those of 
small total energy are darker (blue). Somewhat different lightning 
and shading of the shape was applied in panel (d) and the contrast has 
been enhanced in order to emphasize the 
appearance of depressions in the shell, enclosed also by dotted lines.
}
\label{fig:fig7}
\end{figure}

Figure \ref{fig:fig6} displays how the mean square aspherity changes during the 
buckling transition for the constrained and unconstrained shapes. When 
the constrained volume is larger than $V(\gamma \rightarrow 0)$ [panel (a) 
of Fig. \ref{fig:fig6}], the 
final aspherities that are reached through the transition are notably 
lower in shapes with the fixed enclosed volume constraint, the more so the 
larger the fixed volume. The ridge sharpening transition 
\cite{Lobkovsky1,Lobkovsky2} seems to be suppressed, 
at least within the range of parameters studied here. When the 
constrained volume is smaller than $V(\gamma \rightarrow 0)$ [panel (b) 
of Fig. \ref{fig:fig6}], the shell shape changes more dramatically in 
the buckling transition, which can be seen by extremely large 
aspherities that are characteristic of shapes constrained 
to $V = V_0 = 0.91 V(\gamma \rightarrow 0)$. The aspherities reached 
by these shapes cannot be explained by the flattening of the 
faces of the icosahedron, since the mean square aspherity of the 
perfect icosahedron is 0.0025983.  A clue for such large aspherities 
can be found in Fig. \ref{fig:fig7} that shows buckled shapes for 
large FvK numbers. These 
shapes correspond to points denoted by A, B, C and D in Figs. \ref{fig:fig5} and 
\ref{fig:fig6}, i.e. they are calculated for the unconstrained case (A), in 
case when volumes are constrained to $V = V_0 = V(\gamma \rightarrow 0)$ (B), 
$V = V_0 = 1.1 V(\gamma \rightarrow 0)$ (C) and $V = V_0 = 0.91 V(\gamma \rightarrow 0)$ 
(D). The triangular faces in the shapes 
are colored according to the total energy that they contain. This was calculated 
as one half of the energy contained in the three edges of a particular triangle. 
Note how the distribution of energy contained in the shell changes depending on 
whether the enclosed volume conservation constraint is imposed or not. For 
unconstrained shells, the largest energy is contained in the vicinity of 
the pentagonal disclinations (similar effect was observed 
in Ref. [\onlinecite{Zandistress}]), while the {\em opposite} is observed for the 
shapes in which the constrained volume is larger than $V(\gamma \rightarrow 0)$ 
- in that case the largest energy is contained 
in the regions {\em between} the disclinations that ''bulge out'' 
to satisfy the volume conservation constraint [see panel (c) 
of Fig. \ref{fig:fig7}]. Practically all of this energy is of the stretching type, 
since the distances between the points on the shell (vertices) have to be larger 
from their equilibrium values in order to satisfy the constraint. This may have some 
implications to bursting 
of viral capsids, in particular the points at which the cracks in the 
capsid initiate \cite{Zandistress}. Particularly interesting is the shape obtained 
for $V = V_0 = 0.91 V(\gamma \rightarrow 0)$ [panel (d) of Fig. \ref{fig:fig7}]. 
Some shading is applied in the depiction of this shape [unlike in shapes 
displayed in panels (a),(b), and (c)] in order to clearly emphasize 
its main characteristic - a {\em depression}, collapse, or {\em crumpling} 
of the shell localized in between the disclinations. This is the reason for the 
extremely large aspherities observed in panel (b) of Fig. \ref{fig:fig6}. Note 
that the highest energies in this case are localized in vicinity of the pentagonal 
disclinations, since the shell collapse does not require a large 
change in the stretching energy. 

\begin{figure}[ht]
\centerline{
\epsfig {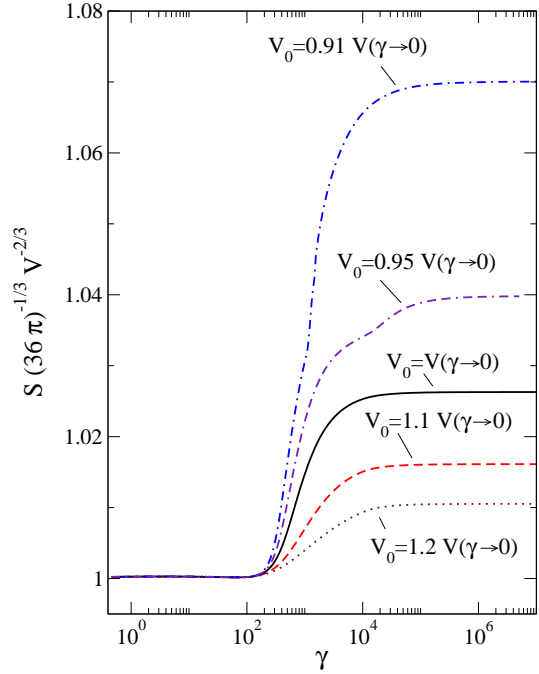}
}
\caption{(Color online) The normalized area of shells with constrained volumes as 
denoted in the figure as a function of the F\"{o}ppl- von K\'{a}rm\'{a}n number 
($T=400$).}
\label{fig:fig7a}
\end{figure}

The total area ($S$) of the shapes also increases during the buckling transition 
under the constraint of fixed volume. The normalized area, $S (36 \pi)^{-1/3} V^{-2/3}$ as 
a function of the FvK number is shown in Fig. \ref{fig:fig7a}. Note that 
for a perfect sphere, the normalized area as defined here is 1, and for 
icosahedron, it is 1.0646. 

\section{Buckling transition in icosahedral shells with the constant internal pressure}
\label{sec:press}

Before studying the effects of pressure on the icosahedral shells, it should be 
decided in what effective units the pressure is to be measured. The quantity 
that was kept fixed in the calculations presented thus far was $\epsilon$, 
that is the scale of energy related to stretching. The internal pressure 
necessarily induced stretching of the shell, and thus, the pressure should 
be expressed in units of $\epsilon$ divided by some spatial scale. It seems 
reasonable to chose the radius of the shell in the limit when 
$\kappa \rightarrow \infty, \gamma \rightarrow 0$ {\em and} $p=0$ as the 
relevant spatial scale. 
The radius of the shell in this limit is denoted as $R_0$, and the pressure is 
thus measured in the units of $\epsilon / R_0$. I shall demonstrate that in 
the continuum limit of shells this indeed produces appropriate results which 
coincide for the shells of different (but still large) $T$ numbers.

\begin{figure}[ht]
\centerline{
\epsfig {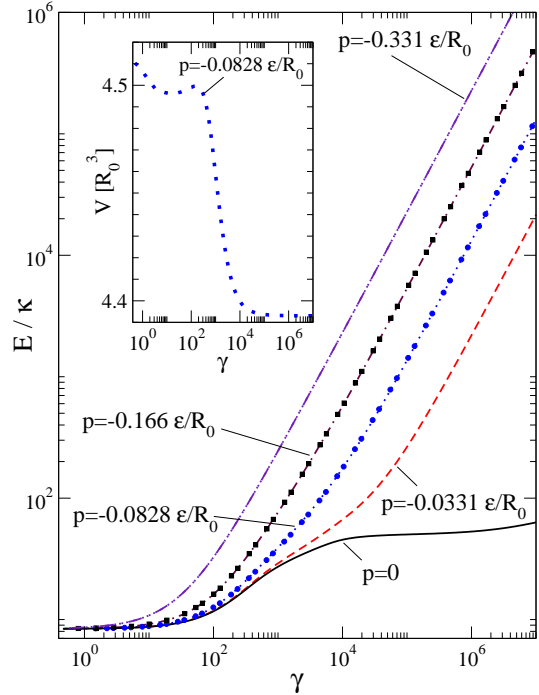}
}
\caption{(Color online) The shell energy as a function of the 
F\"{o}ppl- von K\'{a}rm\'{a}n number for five different values of 
pressure, as denoted in the figure, that is acting on the shell 
from its inside. The calculations represented by the lines were 
performed for $T=400$ shells, and squares and circles represent 
the results of the calculation for $T=625$ and $T=1225$ shells, 
respectively. The shell energies without applied pressure 
($p=0$) are shown for comparison. The inset displays the 
change in volume (measured in $R_{0}^{3}$ units, see text) 
of the $T=400$ shell subjected to 
the internal pressure of $p=-0.0828 \epsilon / R_0$ as 
a function of the F\"{o}ppl- von K\'{a}rm\'{a}n number.
}
\label{fig:fig7b}
\end{figure}

The effects of pressure that acts from the inside of the shell 
(negative pressures) on the 
shell energy is shown in Fig. \ref{fig:fig7b}. This can also be thought 
of as a situation in which the internal pressure in the shell is larger 
from the surrounding pressure. The quantity shown on the 
ordinate axis of Fig. \ref{fig:fig7b} is the shell energy, that is the 
value obtained when $pV$ is subtracted from the total Hamiltonian 
[see Eq. \ref{eq:tlak}]. Similar trends are observed here as in 
panel (a) of Fig. \ref{fig:fig5}. For large internal pressures, the 
shell energies increase significantly above the values that are 
obtained when there is no pressure acting on the shell. The full 
lines in the figure were calculated for $T=400$ shells. The 
radius of this shell in the limit when 
$\kappa \rightarrow \infty, \gamma \rightarrow 0$ is $R_0 = 16.5682 a$ 
[see Eq.(\ref{eq:h0})], so the pressures shown in Fig. \ref{fig:fig7b} 
can also be expressed as -0.002 $\epsilon / a$ (-0.0331 $\epsilon / R_0$), 
-0.005 $\epsilon / a$ (-0.0828 $\epsilon / R_0$), 
-0.01 $\epsilon / a$ (-0.166 $\epsilon / R_0$), and 
-0.02 $\epsilon / a$ (-0.331 $\epsilon / R_0$). The circles 
in Fig. \ref{fig:fig7b} represent the results obtained for the 
$T=625$ shell with the applied pressure $p=-0.008003 \epsilon / a = 
-0.166 \epsilon / R_0$, since $R_0 = 20.7035 a$ for $T=625$ shell. Note 
that these results coincide with those obtained for $T=400$ shell, which 
confirms both that the continuum limit is reached and the 
appropriateness of the units chosen for pressure. The same 
holds for $T=1225$ shell [squares in Fig. \ref{fig:fig7b}] for 
which $R_0 = 28.9738 a$. When 
the constant internal pressure is applied to the shell, as the 
FvK number changes, so does the volume enclosed by the shell. The 
inset in Fig. \ref{fig:fig7b} displays the change of the 
enclosed volume (measured in $R_0^3$ units) with the FvK number 
for $p=-0.0828 \epsilon / R_0$ 
($T=400$). Note that the enclosed volume in the limit 
$\kappa \rightarrow \infty, \gamma \rightarrow 0$ is {\em not} 
$4 \pi / 3 \approx 4.189$ but larger (about 4.6) which is simply 
due to the fact that the effective internal force (pressure) acts 
on the shell and increases its volume. For all the pressures studied, 
the volume of the buckled shapes (in the limit when $\gamma \rightarrow \infty$) is 
{\em smaller} from the volume of unbuckled shapes 
(in the limit when $\gamma \ll \gamma_b$), and the difference between the 
two values of the volume becomes smaller with the increase (in absolute 
value) of the internal 
pressure. Note that the internal pressure necessarily induces the increase 
of the {\em stretching} energy of the shell. The effects of positive 
(outside) pressure on the shell energies and shapes could not be 
studied with the method presented in subsection \ref{sub:tlakham}, since 
for some critical value of the FvK number, depending on 
the magnitude of the applied pressure, the shell abruptly 
collapses, crumples and the numerical method chosen becomes 
unreliable for tracing this effect.

\begin{figure}[ht]
\centerline{
\epsfig {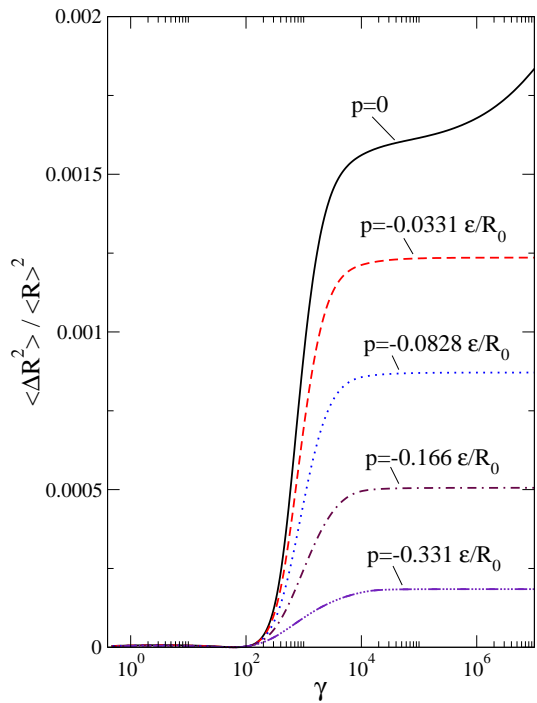}
}
\caption{(Color online) Mean square aspherities of the shells 
subjected to constant internal pressures, as denoted in 
the graph, as a function of the F\"{o}ppl- von K\'{a}rm\'{a}n number 
($T=400$).
}
\label{fig:fig7c}
\end{figure}

Mean square aspherities of the shapes are shown in Fig. 
\ref{fig:fig7c}. It is clear that large internal pressures 
suppress the change in aspherity in the buckling transition, but 
the transition is still observed within the region of pressures studied 
here. Note that this is not a trivial finding since for 
$p=-0.331 \epsilon / R_0$, the enclosed shell volume for 
$\gamma = 0.6$ is 1.4 times larger from the volume that 
would be enclosed by the shell without the applied pressure, yet 
the buckling transition still survives.

\section{Application of the results to viral shapes}
\label{sec:appli}

It is of use to scale the pressures studied in Sec. \ref{sec:press} 
to circumstances that are relevant to viruses. To do this, an estimate of the elastic 
parameters of the viral coatings is needed. The authors of 
Ref. \onlinecite{elastic1} have 
suggested that the ratio $Y / \kappa$ appropriate for 
viruses is about 1-2 nm$^{-2}$. The authors of Ref. 
\onlinecite{elastic2} have taken a step further and they estimate 
that the Young modulus of the viral coating is about $Y = 10 k_B T$/nm$^2$. 
The scale of the stretching energy, $\epsilon$ is of the same order 
of magnitude as $Y$ [see Eq.(\ref{eq:h0}) and the discussion following 
it], thus I shall fix $\epsilon$ to $10 k_B T / nm^2$. The radius 
of the virus in its spherical (immature) form depends on the virus 
in question. If one considers HK97 bacteriophage procapsid of radius 
$R_0 = 26$ nm, the scale of pressure for this virus is 
$\epsilon / R_0 = 0.2 k_B T / nm^3 = 1657 000$ Pa $\approx$ 16.4 atm. 
The highest pressure studied (-0.331 $\epsilon / R_0$) would thus in 
case of HK97 bacteriophage be about 548600 Pa $\approx$ 5.42 atm. 
It is of interest to compare these numbers to those obtained in 
Ref. [\onlinecite{tzlil}]. In studying the energetics of 
DNA inserted in the preformed viral capsids, the authors have 
found \cite{tzlil} that for a virus of radius 27.5 nm, the pressure that 
the {\em fully} packed DNA genome (whose length is 16.5 $\mu$m) exerts 
on its walls is about 30-45 atm. For a virus of such radius, 
the pressure scale introduced in this article should be 
about $\epsilon / R_0 \approx$ 15.5 atm, assuming the same value 
for $\epsilon$ as for the HK97 bacteriophage. The highest pressure 
studied in this work would then correspond to 5.1 atm, about 6 times 
smaller from the pressure acting in the filled viral capsid of 
the virus studied in Ref. \onlinecite{tzlil}. Similar estimate for 
the pressure that the fully loaded genetic material 
exerts on the viral coating was obtained in Ref. \onlinecite{purohit2003}. 
For $\phi$29 bacteriophage, the authors estimate that the internal 
pressure is about 60 atm, in agreement with previous estimates 
from Ref. \onlinecite{NatSmith}. There is an obvious agreement in 
the literature concerning the internal pressures in mature viral capsids. 
However, for such large internal pressures, 
{\em and assuming that} $\epsilon = 10 k_B T$/nm$^2$
the mean square aspherities of buckled shapes (i.e. mature viruses) should 
be very small (see Fig. \ref{fig:fig7c}), far smaller than observed e.g. for 
mature HK97 capsid (0.00116). On the other hand, one could insist on 
the application of the present model to viruses and use 
the results presented here to re-estimate the value of the 
Young modulus of the coating of a mature virus. In other words, 
the aim is to scale the internal pressures to the values obtained in 
previous studies and to obtain new estimates for elastic parameters of 
mature viral capsid. I shall again concentrate on HK97 virus. 
Assuming that the capsid's elastic parameters are in the 
region where the mean square aspherity saturates, the internal 
pressure should be about 0.045 $\epsilon / R_0$ 
(see Fig. \ref{fig:fig7c} - this internal pressure results in final 
aspherity that is about the same as in mature HK97 virus). 
Equating this with the estimated internal 
pressure of about 40 atm \cite{purohit3}, and using $R_0 = 26$ nm, one obtains 
that for the mature HK97 virus, $Y \approx 650 k_B T$/nm$^2$ (2.7 N/m), 
about sixty times larger from the 
estimate of Ref. \onlinecite{elastic2}, but closer to the 
alternative (discarded) estimate of Ref. [\onlinecite{elastic2}] which 
was about $Y \approx $1 N/m. The bending rigidity should then be smaller 
than about 1.8 10$^{-19}$ J ($\gamma > 10^4$, see Fig. \ref{fig:fig7c}). 
Division of the two-dimensional Young's modulus 
obtained in this study with the thickness of the mature HK97 viral shell 
(about 2.5 nm \cite{Conway,Jiang}) yields an approximate prediction for the 3D 
(bulk) Young's modulus ($Y_{3D}$) of the viral protein shell. 
This procedure gives $Y_{3D}$ = 1.1 GPa which is about fourty percent smaller from 
the value found in the experimental study of the elastic response of 
empty viral shells of $\phi$29 bacteriophage with the use of AFM 
microscope (1.8 GPa) \cite{AFMpress}. Note, however, that the results obtained 
here relate to elastic properties of {\em mature} (filled) viral shells, 
while the experimental results are performed on empty capsids. Nevertheless, 
the results obtained here suggest 
that the Young's modulus of the viral coating may be significantly 
larger than predicted in Ref. [\onlinecite{elastic2}] and closer to the 
value discarded by authors of that work. 
The alternative value of Young's modulus was 
discarded by authors of Ref. [\onlinecite{elastic2}] on the 
basis of Monte Carlo simulations of a coarse grained capsomer \cite{coarse}
model which yielded that the bending rigidity $\kappa$ should be 
of the order of the capsomer-capsomer cohesive binding energy. Thus, 
the results presented here may point towards improvements in the 
coarse-grained model of viruses.

Pouring a liquid into a spherical elastic shell 
(immature capsid), or 
applying a pressure to it from the inside will not, 
of course, change its shape to a more aspherical one. What ought 
to happen "during pouring" (insertion of a genetic material) is a 
{\em dynamical process that changes the elastic properties of a capsid}, i.e. that 
effectively increases its FvK number. Only then 
will a mature capsid be more aspherical than the precursor 
capsid. So, the results presented in this article cannot, of course, provide 
the full physical description of virus maturation. Nevertheless, if the 
change of the capsid shape is influenced by the fact that the capsid is 
filled with the genetic material (in the most trivial sense of finite 
volume occupation or simulated by the constant internal pressure), the results 
should prove useful. The 
presented results essentially show that {\em if} there is a process that 
increases the effective FvK number of the capsid, 
the volume conservation constraint or finite internal pressure 
that may be imposed by the inserted genetic material is not strong enough 
to destroy the signatures of the buckling transition, and the ''mature'' shell 
will be more aspherical from its ''precursor'' shell, although to 
lesser amount than could be concluded from the study of buckling transition 
in {\em empty} shells (see Fig. \ref{fig:fig6}). Now, what could be the 
process that induces the change of elastic properties of the viral capsid?

\begin{figure}[ht]
\centerline{
\epsfig {file=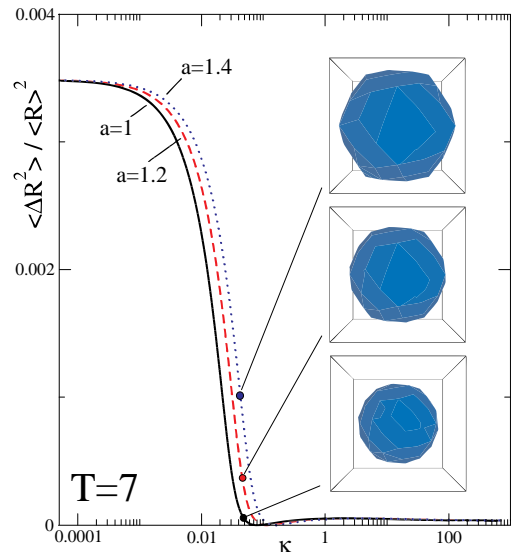,width=7cm}
}
\caption{(Color online) Mean square aspherities of $T=7$ shells as a function of bending 
modulus. The calculations for three different values of the lattice constant 
$a$ are displayed ($a=1.0$ - full line, $a=1.2$ - dashed line, and 
$a=1.4$ - dotted line). The three shapes characterized by these three 
lattice constants are plotted for $\kappa=0.0458$.}
\label{fig:fig8}
\end{figure}

The proposition put forth by Lidmar {\em et al} \cite{elastic1} is 
that the FvK number changes during the maturation 
due to the {\em increase} of radius ($R$) and {\em decrease} of effective thickness 
of the protein coating, $d$. According to Ref. [\onlinecite{elastic1}], 
if one approximates a shell by a uniform 
isotropic elastic medium of thickness $d$ with the Poisson ratio 
$\nu_3$, then its effective FvK number should be about 
\begin{equation}
\gamma = 12 (1-\nu_3^2)(R/d)^2.
\end{equation}
If one additionally assumes that the volume of the capsid protein 
material (not the volume {\em 
enclosed} by the capsid) remains the same, and that $\nu_3$ 
does not change during maturation, then the 
FvK numbers before ($\gamma_B$) and after ($\gamma_A$) the 
transition for the HK97 capsid studied in Ref. \onlinecite{elastic1} should be 
$\gamma_A = 1480$ and $\gamma_B \approx 630$. This indeed predicts a noticeably more 
spherical shape of the immature capsid. It has been experimentally observed 
that the capsid 
proteins rearrange themselves significantly during maturation \cite{Steven}. To a lesser 
extent, the reorganization involves the refolding of the protein itself, but its  
main consequence is a rearrangement (translations and rotations) of the proteins 
to another configuration in which the contacts between capsomers change and 
the capsomers adopt a more symmetrical shape which also results in a 
smoother shape of the capsid. This rearrangement has been 
studied in detail for bacteriophage HK97 and P22 capsid in Refs. [\onlinecite{Conway}] and 
[\onlinecite{Jiang}], respectively. These experimental findings suggest 
that one should study the 
effects of a change of the lattice constant, $a$ on the shape of the 
minimum energy shells. This was done in Fig. \ref{fig:fig8}. The calculations were 
intentionally performed for small value of $T$-number ($T=7$, which corresponds to 
the symmetry of HK97 capsid), although one should keep in mind that the Hamiltonian in 
Eq. (\ref{eq:h0}) was constructed with the idea of treating the continuous (large 
$T$-number) structures for which it makes sense to speak about elastic parameters 
of protein sheets, and its predictions for small $T$-numbers should not be 
directly transcribed to viral shapes. Nevertheless, the form in which the 
Hamiltonian is written [Eq. (\ref{eq:h0})] suggests its ''microscopic'' 
interpretation. These subtleties were discussed in Refs. 
[\onlinecite{elastic1,elastic2}]. The calculations in Fig. \ref{fig:fig8} were 
performed for three values of the lattice constant, $a=1, a=1.2, a=1.4$, scale 
of the elastic energy was kept constant ($\epsilon=1$), and the bending 
modulus $\kappa$ was changed. The mean square aspherities were plotted as 
a function of $\kappa$. The constraint of fixed enclosed volume was 
not implemented and $p=0$.  Note that in the ''critical'' region of values of 
$\kappa$ (onset of the buckling transition), the 
shapes with very different aspherities can be produced by simply changing the 
lattice parameter $a$, and keeping all other parameters of the Hamiltonian 
fixed (for the shapes plotted in Fig. \ref{fig:fig8}, the bending modulus 
was $\kappa = 0.0458$). This numerical experiment reproduces the experimental findings that 
the $i)$ mature viral shapes enclose larger volume, and $ii)$ mature viral 
shapes are more facetted from their precursor shapes. Of course, it 
still does not provide any clues with respect to the process that 
drives the transition to larger lattice constant.

\section{Summary and Conclusions}
\label{sec:discuss}
The buckling transition of shells with twelve pentagonal disclinations situated in 
vertices of an icosahedron was studied. The constraint of the 
fixed enclosed volume was introduced and its effect on the shell shapes and 
energies explored. The shell shapes and buckling 
transition were also investigated in cases when the constant internal 
pressure acts on the shell. It was found that the buckling transition survives the 
volume constraint, at least for the enclosed volumes that are not too large. The 
buckled shapes are found to be less aspherical than the ones obtained without 
the volume constraint. Similar conclusions were reached for shells subjected 
to constant internal pressure - for pressures that are not too large 
(about $p = - 0.3 \epsilon / R_0$), the buckling transition survives and 
the buckled shapes are significantly more aspherical from the ones 
in the region of small FvK numbers. If one transcribes the results of 
the study of shells under constant internal pressures to the 
circumstances appropriate to viruses \cite{purohit2003,NatSmith}, an 
estimate of the two-dimensional Young's modulus of the coating can be obtained, and 
for mature HK97 virus, the value of $Y \approx 2.7$ N/m was found. This 
also puts an upper limit on bending rigidity, $\kappa \le$ 1.8 10$^{-19}$ J.
For the bulk (3D) Young's modulus of the viral coating material this yields a 
value of about $Y_{3D} \approx$  1.1 GPa, which is close to the 
value found in experimental studies of elastic response of empty  
$\phi$29 bacteriophage shells \cite{AFMpress}.

Whether the fixed enclosed volume of the shell, 
or the constant internal pressure 
is a realistic model of the (mature) viral shells filled with the genetic material 
is, of course, questionable and alternative scenarios for the virus maturation, 
based again on the triangulated elastic shells, have been proposed in 
Sec. \ref{sec:appli}. In any case, the results of this article should be 
of use in extending the approaches of Refs. [\onlinecite{elastic1,elastic2}] 
towards more realistic modeling of viral shapes.


\begin{thebibliography}{}
\bibitem{primjedba} I use the definition of F\"{o}ppl-von K\'{a}rm\'{a}n number from 
Ref. \onlinecite{elastic1} which differs from the one in Ref. \onlinecite{elastic2} 
by a multiplicative factor of $(4 \pi)^{-1}$.
\end{thebibliography}
\end{document}